\documentclass[twocolumn,showpacs,preprintnumbers,nofootinbib,superscriptaddress]{revtex4}

\usepackage{graphicx}
\usepackage{latexsym}
\usepackage{amssymb}
\usepackage{amsmath}
\usepackage{stmaryrd}
\usepackage[british]{babel}
\usepackage[none]{hyphenat}

\DeclareMathOperator{\arctanh}{arctanh}

\begin{document}

\title{A Quantum Field Theory View of Interaction Free Measurements}

\author{Filipe Barroso}
\email{filipe.barroso@fc.up.pt}
\affiliation{Departamento de Física e Astronomia, Faculdade de Ciências da Universidade do Porto, and}
\author{Orfeu Bertolami}
\email{orfeu.bertolami@fc.up.pt}
\affiliation{Departamento de Física e Astronomia, Faculdade de Ciências da Universidade do Porto, and}
\affiliation{Centro de Física do Porto, Faculdade de Ciências da Universidade do Porto, both at Rua do Campo Alegre 1021/1055, 4169-007 Porto, Portugal}

\vskip 0.5cm

\date{\today}

\begin{abstract}

We propose a Quantum Field Theory description of beams on a Mach-Zehnder Interferometer and apply the method to describe Interaction Free Measurements (IFMs), concluding that there is a change of momentum of the fields in IFMs. Analysing the factors involved in the probability of emission of low-energy photons, we argue that they do not yield meaningful contributions to the probabilities of the IFMs.

\vskip 0.5cm

\end{abstract}

\maketitle
\vskip 2pc
\section{Introduction}

Avshalom C. Elitzur and Lev Vaidman proposed, in 1993, a method to obtain information about the presence of an object without interacting with it, a procedure dubbed as Interaction Free Measurement (IFM) \cite{EV}. The key issue of IFMs is the superposition of quantum mechanical (QM) states.

\begin{figure}[h]
\includegraphics[scale=0.385]{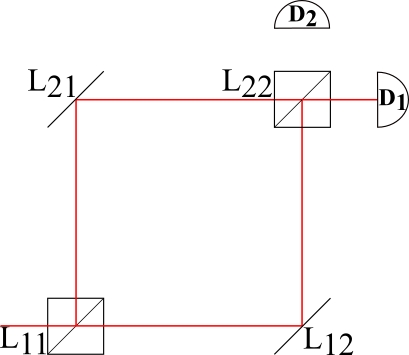}\includegraphics[scale=0.385]{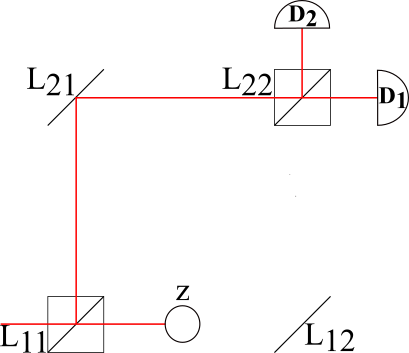}
  \caption{MZI with a bomb (right) and without it (left). In the absence of the bomb, only $D_1$ detects photons. When the bomb, in $z$, blocks the way of one of the beams, both detectors tick. $L_{ij}$ are labels for the vertices of the MZI. \label{fig:interferometer}}
\end{figure}

The method, in its simplest form, considers a photon beam in a Mach\--Zehnder interferometer (MZI) in which a bomb is placed on one of its arms (see Fig.~\ref{fig:interferometer}). Beams are assumed to be completely reflected by the mirrors ($\diagup{}$), while the beamsplitters ($\boxslash{}$) reflect half the beam and transmit the rest. Without loss of generality, it can be assumed that an incoming particle on the boobytrapped path will always interact with and trigger the bomb. This interaction is local and the arms of the interferometer are sufficiently far apart so that particles in the other path would not interact with the bomb. Without the bomb, every particle in the interferometer would be detected by single detector, say $D_1$; with the bomb, both detectors share a quarter of probability of detection, with a further half being the probability of triggering the bomb.

The conclusion of this experiment is that, whenever the second detector, $D_2$, ticks, there is an object blocking one of the paths of the interferometer. Note that the experiment can be executed with any opaque object, but its usefulness derives from the fact that it allows for inferring the existence of unstable states without interacting with them. Such states are modelled by the bomb and its propensity to detonate once it interacts. More complex configurations can improve the rate of detection of the bomb without exploding it, assuming it is placed inside a cavity, possibly up to the theoretical limit of $100\%$ \cite{Paul,Kwiat,Namekata}, but the simpler scheme discussed above suffices to make a meaningful analysis.

In this work, we shall describe the IFMs from the point of view of Quantum Field Theory (QFT). This formalism presents a more encompassing way for discussing interaction ranges, scattering and localisation of particles. This formalism is also appropriate when discussing Weinberg's Soft Photons Theorem \cite{Bloch,Weinberg, Weinberg2} and its implications, which we shall use to discuss the emission of photons with low energy.

The possibility of having an infinitely sensitive bomb has raised some criticism and discussion (see Refs. \cite{Simon,Vaidman}). Indeed, an infinitely sensitive bomb would be physically unreasonable: quantum fluctuations or the emission of the infinite amount of soft photons could trigger the bomb.\footnote{Notice that if the detonation is detectable, these photons cannot be soft as soft photons have, by definition, such low energy that they evade detection. To avoid this semantic confusion, we shall refer to the detectable photons as low-energy photons.}~In Section~\ref{sec:softphotons}, we shall discuss the impossibility of having a workable experiment in this regime due to low-energy photons. On the other hand, we show that if the detection limits are nonvanishing, the emission of low-energy photons can be ignored. Another point of contention regards the transference of momentum \cite{Simon,Karlsson}: it is argued that energy is unchanged in IFMs, but momentum is not. In our QFT description of IFMs, in Section~\ref{sec:QFTD}, we show that the field detected in $D_2$ has the same energy, but a rotated momentum with respect to the initial field.

\section{Quantum field theory description}\label{sec:QFTD}

Given the suggested sensibility of the bomb, it will be considered as a quantum mechanical object. In order to introduce more naturally the concept of interaction, we shall make use of the QFT description. A standard derivation can be found, for instance, in Ref. \cite{EV}.

In our analysis, we assume two interacting fields: one for the beam in the interferometer and another for the bomb. To accommodate the discussion of Section~\ref{sec:softphotons}, we shall use photonic and fermionic fields, respectively, however a simpler derivation for scalar fields could be considered instead. The mirrors and beamsplitters, assumed to be perfect, will be defined as unitary transformations that act only on a set of positions of the field. To simplify the discussion, let us assume that those devices are actually fixed in a single point in space, at any in time. The beamsplitter divides the field in two components that can be regarded as independent of each other if they are well localised perturbations. Propagating these components through the mirrors and beamsplitters in the order described above reproduces the resulting interferometry. The effect of the bomb is simulated through an interaction in one, and only one, of the paths. We assume the interaction to be with a fermion in the lower path.

\subsection{Fields and Unitary Operators}
Consider the photon field, expanded in Fourier modes with energy $E_\mathbf{p}$ and polarisations $\epsilon^{*\lambda}_\mu(\mathbf{p})$, $\epsilon^\lambda_\mu(\mathbf{p})$ \cite{Peskin},
\begin{equation}
\begin{split}
    A_\mu(x)=\sum_\lambda\int{\frac{d^3 p}{\left(2\pi\right)^3 2E_\mathbf{p}}}& \bigg(\epsilon^\lambda_\mu(\mathbf{p}) a_\lambda(\mathbf{p})e^{ip \cdot x} + \\ & +\epsilon^{*\lambda}_\mu(\mathbf{p})a^\dagger_\lambda(\mathbf{p})e^{-ip \cdot x}\bigg),
\end{split}
\end{equation}
subject to equal-time commutation relations with its canonical conjugated momentum, $\pi_\nu(t,\mathbf{y})$:
\begin{equation}
\left[ A_\mu(t, \mathbf{x}) , \pi_\nu(t,\mathbf{y}) \right]=i\eta_{\mu\nu}\delta^{(3)}(\mathbf{x}-\mathbf{y}),
\end{equation}
while the remaining commutation relations vanish.

Furthermore, consider the unitary operator, dependent on a parameter $\alpha$ (see Ref. \cite{GerryKnight}),
\begin{equation}
\begin{split}
    V(\alpha)=\exp{}&\Bigg( i\alpha R^{\mu\nu}\sum_{\lambda,\kappa}\int{\frac{d^3 p}{\left(2\pi\right)^3 2E_\mathbf{p}}}\cdot \\& \cdot\epsilon^{*\lambda}_\mu(\mathbf{R}\cdot\mathbf{p}) a^\dagger_\lambda(\mathbf{R}\cdot\mathbf{p})a_\kappa(\mathbf{p})\epsilon^{\kappa}_\nu(\mathbf{p})\Bigg),
\end{split}
\end{equation}
where $R^{\mu\nu}$ is a unitary transformation in Minkowski space, with $R^\nu_\mu R^\alpha_\nu=\delta^\alpha_\mu$, and where $\mathbf{R}$ is its purely spatial part. In particular, to simulate reflections, we require $\mathbf{R}$ to be an Householder matrix \cite{Cabrera} and $R^{\mu\nu}$ its extension to space-time is such that it is independent of time component:
\begin{equation}
R^{\mu\nu}=\begin{pmatrix}1 & 0 \\ 0 & \mathbf{R} \end{pmatrix}
\end{equation}
and
\begin{equation}
\mathbf{R}=\mathbf{1}-2\hat{\mathbf{n}}\otimes\hat{\mathbf{n}},
\end{equation}
where $\hat{\mathbf{n}}$ is defined as the unit vector normal to the surface of the mirror. Hence, $A^\prime_\mu(R\cdot x)$ has a momentum reflected with respect to the one from $A_\mu(x)$, while the field energy remains unchanged.\footnote{The 4-momentum operator is
\begin{equation}
P^\mu=\int{T^{0\mu}d^3x},
\end{equation}
where $T^{\nu\mu}$ is the energy-momentum tensor \cite{Peskin}.}~Specific values for the angle $\alpha$ model the actions of the mirrors and beamsplitters. The equivalent operator for scalar beams does not involve $R^{\mu\nu}$ nor the polarisations.

Using the Baker-Hausdorff Lemma \cite{GerryKnight} to expand the expressions for $A_\mu(x)$ and collecting even and odd terms, we get the transformation of the field under $V(\alpha)$:
\begin{equation}
V^\dagger(\alpha)A_\mu(x)V(\alpha)=\cos{(\alpha)}A_\mu(x)+\sin{(\alpha)}A^\prime_\mu(R\cdot x),
\end{equation}
where $A^\prime_\mu(R\cdot x)=V^\dagger(\pi/2)A_\mu(x)V(\pi/2)$ is the part of the transformed field that is orthogonal to the initial field. Similarly, $A^\prime_\mu(R\cdot x)$ transforms as
\begin{equation}
V^\dagger(\alpha)A^\prime_\mu(R\cdot x)V(\alpha)=\cos{(\alpha)}A^\prime_\mu(R\cdot x)-\sin{(\alpha)}A_\mu(x).
\end{equation}

Due to the unitarity of $V(\alpha)$, the commutation relations at equal-time for the transformations of $A_\mu(x)$ and $\pi_\mu(x)$ remains unchanged:
\begin{equation}\begin{split}
\big[ V^\dagger(\alpha) A_\mu(t,\mathbf{x}) V(\alpha) ,& V^\dagger(\alpha) \pi_\nu(t,\mathbf{y}) V(\alpha) \big]= \\ &= i\eta_{\mu\nu}\delta^{(3)}(\mathbf{x}-\mathbf{y}).
\end{split}\end{equation}
In particular, for $\alpha=\pi/2$, we have the equal-time commutation relation
\begin{equation}
\left[ A^\prime_\mu(t,\mathbf{x}) , \pi^\prime_\nu(t,\mathbf{y}) \right]=i\eta_{\mu\nu}\delta^{(3)}(\mathbf{x}-\mathbf{y}).
\end{equation}

\subsection{Mirrors and Beamsplitters}
The mirror is defined as an operator that acts on the vector field through $V(\pi/2)$ in a spatial point $\mathbf{z}$, for any value of time, and is an identity otherwise. Explicitly, this action yields:
\begin{equation}
\diagup{}_z\left[ A_\mu(x) \right]= \begin{cases} A^\prime_\mu(R\cdot x) & \mbox{if } \mathbf{x}=\mathbf{z}, \\ A_\mu(x) & \mbox{otherwise}. \end{cases}
\end{equation}

Similarly, a point-like beamsplitter placed in $\mathbf{z}$ is defined as a unitary transformation on the field under $V(\pi/4)$, subject to the same conditions as the mirror. This leads to a change of the field as
\begin{equation}
\boxslash{}_z\left[ A_\mu(x) \right]= \begin{cases} \frac{1}{\sqrt{2}}\left(A_\mu(x) + A^\prime_\mu(R\cdot x)\right) & \mbox{if } \mathbf{x}=\mathbf{z}, \\ A_\mu(x) & \mbox{otherwise}. \end{cases}
\end{equation}

\subsection{Localised Perturbations and the Interferometer}
The computation of the equal-time commutation relations, through the expansion in Fourier modes for the components $A_\mu(x)$, $A^\prime_\mu(x)$ and their conjugated momenta, yields
\begin{equation} \label{eq:lastline}
\left[ A_\mu(t,\mathbf{x}) , \pi^\prime_\nu(t,\mathbf{y}) \right]=\left[ A^\prime_\mu(t,\mathbf{x}), \pi_\nu(t,\mathbf{y}) \right]=i\eta_{\mu\nu}\delta(\mathbf{x}-\mathbf{y}).
\end{equation}
If the fields are well localised and their tails can be disregarded, the commutators in Eq. (\ref{eq:lastline}) vanish,
\begin{equation}\begin{split}
\left[ A_\mu(t,\mathbf{x}) , \pi^\prime_\nu(t,\mathbf{y}) \right]=\left[ A^\prime_\mu(t,\mathbf{x}), \pi_\nu(t,\mathbf{y}) \right]\approx0,
\end{split}\end{equation}
allowing for treating the field and its reflected counterpart as independent. Thus, each component can be propagated independently.

To define the interferometer, we consider four points in space-time: $L_{11}$, $L_{12}$, $L_{21}$ and $L_{22}$ (see Fig.~\ref{fig:interferometer}); we place the beamsplitters in the first and final points and place mirrors in the remaining pair. Destructive interference occurs when the fields are propagated through these points on the following order: splitting of a field $A_\mu(x)$ in $L_{11}$, propagation of $\frac{1}{\sqrt{2}}A_\mu(L_{11})$ to $L_{21}$ and $\frac{1}{\sqrt{2}}A^\prime_\mu(R\cdot L_{11})$ to $L_{12}$, reflection of both components on those points and, finally, propagation of both of them to the same point $L_{22}$, where a beamsplitter operates. Representing schematically these transformations corresponds to
\begin{equation}\begin{split}
A&_\mu(L_{11}) \xrightarrow[]{\boxslash{}_{L_{11}}} \frac{1}{\sqrt{2}}\left(A_\mu(L_{11})+A^\prime_\mu(R\cdot L_{11})\right) \xrightarrow[\diagup{}_{L_{21}}]{\diagup{}_{L_{12}}} \\& \frac{1}{\sqrt{2}}\left(A^\prime_\mu(R\cdot L_{12})-A_\mu(L_{21})\right) \xrightarrow[]{\boxslash{}_{L_{22}}} -A_\mu(L_{22}),
\end{split}\end{equation}
where propagations are to be assumed between the different points, allowing for the computation of probability amplitudes. The probability amplitude of a path inside the interferometer is written, in terms of two-point correlation functions $G(\cdot,\cdot)$ \cite{Peskin}, as
\begin{equation}
G(L_{11},L_{12})G(L_{11},L_{21})G(L_{12},L_{22})G(L_{21},L_{22}),
\end{equation}
where $G(x,y)=\langle\Omega|T\{A_\mu(x)A_\nu(y)\} |\Omega\rangle$ and $T\{\cdot\}$ is the time-ordering operator.

\subsection{Interaction}
In order to include an interaction, for example, with a fermion, in the path between $L_{11}$ and $L_{12}$, we substitute $G(L_{11},L_{12})$ by
\begin{equation}
G(L_{11},z_1)\langle\Omega| A_\mu(z_1)A^{\prime\prime}_\nu(z_2)\psi(x)\bar{\psi}(y) |\Omega\rangle G(z_2,L_{12}),
\end{equation}
where $z_1$ and $x$ are constrained by the cross section of the interaction, since we require the interaction to occur; $A^{\prime\prime}_\nu(z_2)$ is the photon field after interaction and $\psi(x)$, $\bar{\psi}(y)$ are the initial and final fermion fields, respectively. The term $\langle\Omega|\cdot|\Omega\rangle$ corresponds to a scattering matrix, reproducing the scattered state as described in Ref. \cite{EV}. To reproduce IFMs, the size of the arms of the interferometer must be much greater than the radius of the cross section of the process, such that the interaction only occurs on one of the paths.

The new set of transformations is
\begin{equation}\begin{split}
&A_\mu(L_{11}) \xrightarrow[]{\boxslash{}_{L_{11}}} \frac{1}{\sqrt{2}}\left(A_\mu(L_{11})+A^\prime_\mu(R\cdot L_{11})\right)\xrightarrow{e^{-}}
\\& \frac{1}{\sqrt{2}}\left(A^{\prime\prime}_\mu(z_2)+A^\prime_\mu(R\cdot L_{11})\right)
\xrightarrow[]{\diagup{}_{L_{21}}} \frac{1}{\sqrt{2}}\left(A^{\prime\prime}_\mu(z_2)-A_\mu(L_{21})\right) \\& \xrightarrow[]{\boxslash{}_{L_{22}}} \frac{1}{\sqrt{2}}A^{\prime\prime}_\mu(z_2)-\frac{1}{2}\left(A_\mu(L_{22})+A^\prime_\mu(R\cdot L_{22})\right),
\end{split}\end{equation}
yielding a probability of detection of a reflected field, $A^\prime_\mu(R\cdot x)$, in $D_2$, of $1/4$, reproducing the result obtained by purely QM arguments. Note that the field detected in $D_2$ has the same energy, but a rotated momentum with respect to the initial field.

\section{Pollution by low-energy photons}\label{sec:softphotons}

Considering the bomb as a fermion, we examine in this section the possibility of affecting the standard probabilities associated with the Elitzur and Vaidman's setup due to the emission of low-energy photons. The probabilities cannot be altered by soft photons since thanks to Weinberg's Soft Photons Theorem, they can be renormalised and hence evade detection. We regard low-energy photons as photons subject to the same approximations as soft photons, but detectable.

The emission of $N$ of those photons in a QED process follows a Poisson distribution \cite{Peskin, Weinberg},
\begin{equation}
P\left(N;\mu\right)=\frac{\mu^N}{N!}e^{-\mu},
\end{equation}
where the mean value, $\mu$, is given by
\begin{equation}
\mu=A\left(\alpha\to\beta\right)\ln{\left(\frac{E_{+}}{E_{-}}\right)}.
\end{equation}
The factor $A\left(\alpha\to\beta\right)$ is the Weinberg's factor \cite{Weinberg,Weinberg2},
\begin{equation}
A\left(\alpha\to\beta\right)=-\sum_{n,m}\frac{e_n e_m \eta_n \eta_m}{\left(2\pi\right)^2\beta_{mn}}\arctanh{\left(\beta_{mn}\right)},
\end{equation}
dependent only on the charges, $e_n$, $e_m$, and relative velocities, $\beta_{nm}$ of the particles in the process. The factors $\eta_n$, $\eta_m$ are equal to $\pm 1$, depending on if the particles enter or leave the process. The energies $E_{+}$ and $E_{-}$ are the maximal and minimal energies allowed to the emitted photons, respectively.

The Weinberg's factor, in the case of a fermion scattered by a photon, is given by \cite{Weinberg},
\begin{equation}
A\left(f\to f^\prime\right)=\frac{2 e^2}{\left(2\pi\right)^2}\left[\frac{1}{\beta}\arctanh{\left(\beta\right)-1}\right].
\end{equation}
This expression, despite being divergent for $\beta=1$, for a fermion that leaves the process with a velocity of $99.99\%$ relative to its initial velocity, is only approximately $10e^2$. Hence, this factor does not contribute significantly to $\mu$.

On the other hand, the contribution from $E_{-}$ is logarithmic. If $E_{-}$ is allowed to be nonvanishingly small, similarly to soft photons \cite{Weinberg,Bloch}, $\mu$ diverges, leading to a cloud of low-energy photons emitted in all directions and hitting the detectors. However, if a finite minimal value for detection is assumed, the logarithm will be small, leading to an equally small mean value of emitted photons. In this case, the emission of photons in the direction of the detector is highly unlikely. We thus conclude that the predictions of Elitzur and Vaidman are safe from pollution by these photons, when perfect detectors are discarded.

\section{Conclusions}

We have shown that when considering IFMs in a MZI, the effect of mirrors and beamsplitters can be described by unitary local transformations of quantum fields. The beamsplitters divide well-localised perturbations in two superimposed components that can be treated as independent fields, if allowed to propagate far enough from the beamsplitter. This QFT description of the interferometer allows for a more natural justification of the scattering and finite interaction ranges that are assumed in a somewhat \textit{ad hoc} way in the QM description. When performing IFMs, the momentum of the detected field that signals the existence of the bomb is altered with respect to the initial field. 

We argue that the rate of emission of low-energy photons is small as long as the lower energy threshold for detection is nonvanishing. Only when the emitted photons are allowed to take infinitely low energies, their number might grow to infinity, covering the whole 4-momentum space. In this case, some photons are emitted in the direction of the detector, which fires even after interactions. This means that there will be no sea of low-energy photons if a sensible lower limit detection threshold is assumed, introducing only negligible corrections to the QM probabilities of IFMs.


\end{document}